\documentclass[epj,final]{svjour}
\usepackage{epsfig}

\begin{document}
%\hugehead
\onecolumn
\title{New experimental limits on violations of the Pauli
exclusion principle obtained with the Borexino Counting Test Facility}

\author{
H.O.~Back\inst{a}, M.~Balata\inst{b}, A.~de~Bari\inst{c}, A.~de~Bellefon\inst{d},
G.~Bellini\inst{e}$^{,*}$\thanks{$^{*}$~Spokesman \newline {$^{**}$~Project manager}
\newline {$^{***}$~Operational manager} \newline {$^{\circ}$~GLIMOS} \newline
{$^{\diamond}$~Detector installation manager}\newline  {$^{1}$~{Now at Massachusetts Institute of
Technology, NW17-161, 175 Albany St. Cambridge, MA 02139} }\newline {$^{2}$~Marie Curie fellowship
at LNGS}\newline {$^{3}$~On leave of absence from St. Petersburg Nuclear Physics Inst. - Gatchina,
Russia} \newline {$^{4}$~Now at Sudbury Neutrino Observatory, INCO Creighton Mine, P.O.Box 159
Lively, Ontario, Canada, P3Y 1M3} \newline {$^{5}$~Now at Institute for Nuclear Physics,
Forschungszentrum Karlsruhe, Postfach 3640, 76021 Karlsruhe } \newline {$^{6}$~Now at Institute for
Nuclear Research, Prospekt Nauki 47, MSP 03680, Kiev, Ukraine}\newline {$^{7}$~On leave of absence
from KFKI-RMKI, Konkoly Thege ut 29-33 H-1121 Budapest, Hungary} \newline {$^{8}$~Present Address:
Department of Physics, Virginia Polytechnic Institute and State University, Blacksburg VA  24061
}\newline {$^{9}$~Center for Cosmological Physics, University of Chicago, 933 E. 56$^{th}$St.,
Chicago, IL 60637}\newline {$^{10}$~Deceased} }, J.~Benziger\inst{f}, S.~Bonetti\inst{e},
C.~Buck\inst{g}, B.~Caccianiga\inst{e}, L.~Cadonati\inst{f}$^{,1}$, F.~Calaprice\inst{f},
G.~Cecchet\inst{c}, M.~Chen\inst{h}, A.~Di~Credico\inst{b}, O.~Dadoun\inst{d}$^{,2}$,
D.~D'Angelo\inst{i}, A.~Derbin\inst{j}$^{,3}$, M.~Deutsch\inst{k}$^{,10}$, A.~Etenko\inst{m},
F.~von~Feilitzsch\inst{i}, R.~Fernholz\inst{f}, R.~Ford\inst{f}$^{,4}$, D.~Franco\inst{e},
B.~Freudiger\inst{g}$^{,2,5}$, C.~Galbiati\inst{f}, S.~Gazzana\inst{b}$^{,\circ}$,
M.G.~Giammarchi\inst{e}, M.~Goeger-Neff\inst{i}, A.~Goretti\inst{b}, C.~Grieb\inst{i},
W.~Hampel\inst{g}, E.~Harding\inst{f}, F.X.~Hartmann\inst{g}, G.~Heusser\inst{g}, A.~Ianni\inst{b},
A.M.~Ianni\inst{f}, H.~de~Kerret\inst{d}, J.~Kiko\inst{g}, T.~Kirsten\inst{g},
V.V.~Kobychev\inst{b}$^{,6}$, G.~Korga\inst{e}$^{,7}$, G.~Korschinek\inst{i}, Y.~Kozlov\inst{m},
D.~Kryn\inst{d}, M.~Laubenstein\inst{b}, C.~Lendvai\inst{i}$^{,2}$, M.~Leung\inst{f},
E.~Litvinovich\inst{m}, P.~Lombardi\inst{e}$^{,\diamond}$, I.~Machulin\inst{m},
S.~Malvezzi\inst{e}, J.~Maneira\inst{h}, I.~Manno\inst{o}, D.~Manuzio\inst{n}, G.~Manuzio\inst{n},
F.~Masetti\inst{l}, A.~Martemianov\inst{m}$^{,10}$, U.~Mazzucato\inst{l}, K.~McCarty\inst{f},
E.~Meroni\inst{e}, G.~Mention\inst{d}, L.~Miramonti\inst{e}, M.E.~Monzani\inst{e},
V.~Muratova\inst{j}$^{,3}$, P.~Musico\inst{n}, L.~Niedermeier\inst{i}$^{,2}$, L.~Oberauer\inst{i},
M.~Obolensky\inst{d}, F.~Ortica\inst{l}, M.~Pallavicini\inst{n}, L.~Papp\inst{e}$^{,7}$,
L.~Perasso\inst{e}, P.~Peiffer\inst{g}, A.~Pocar\inst{f}, R.S.~Raghavan\inst{p}$^{,8}$,
G.~Ranucci\inst{e}$^{,**}$, A.~Razeto\inst{n}, A.~Sabelnikov\inst{e}, C.~Salvo\inst{n}$^{,***}$,
R.~Scardaoni\inst{e}, D.~Schimizzi\inst{f}, S.~Schoenert\inst{g}, H.~Simgen\inst{g},
T.~Shutt\inst{f}, M.~Skorokhvatov\inst{m}, O.~Smirnov\inst{j}, A.~Sonnenschein\inst{f}$^{,9}$,
A.~Sotnikov\inst{j}, S.~Sukhotin\inst{m}, Y.~Suvorov\inst{m}, V.~Tarasenkov\inst{m},
R.~Tartaglia\inst{b}, G.~Testera\inst{n}, D.~Vignaud\inst{d}, R.B.~Vogelaar\inst{a},
V.~Vyrodov\inst{m}, M.~Wojcik\inst{q}, O.~Zaimidoroga\inst{j}, G.~Zuzel\inst{q} }

\date{22 June 2004}
\twocolumn
 \institute { {$^{a}${Physics Department, Virginia Polytechnic Institute and State
University, Robeson Hall, Blacksburg, VA 24061-0435, USA}}\\ {$^{b}${I.N.F.N Laboratori Nazionali
del Gran Sasso, SS 17 bis Km 18+910, I-67010 Assergi(AQ), Italy}}\\ {$^{c}${Dipartimento di Fisica
Nucleare e Teorica Universita` and I.N.F.N., Pavia, Via A. Bassi, 6 I-27100, Pavia, Italy}}\\
{$^{d}${Laboratoire de Physique Corpusculaire et Cosmologie, 11 place Marcelin Berthelot 75231
Paris Cedex 05, France}}\\ {$^{e}${Dipartimento di Fisica Universit\`a and I.N.F.N., Milano, Via
Celoria, 16 I-20133 Milano, Italy}}\\ {$^{f}${Dept. of Physics,Princeton University, Jadwin Hall,
Washington Rd, Princeton NJ 08544-0708, USA}}\\ {$^{g}${Max-Planck-Institut fuer
Kernphysik,Postfach 103 980 D-69029, Heidelberg, Germany}}\\ {$^{h}${Dept. of Physics, Queen's
University Stirling Hall, Kingston, Ontario K7L 3N6, Canada}}\\ {$^{i}${Technische Universitaet
Muenchen, James Franck Strasse, E15 D-85747, Garching, Germany}}\\ {$^{j}${Joint Institute for
Nuclear Research, 141980 Dubna, Russia}}\\ {$^{k}${Dept. of Physics Massachusetts Institute of
Technology,Cambridge, MA 02139, USA}}\\ {$^{l}${Dipartimento di Chimica Universit\`a, Perugia, Via
Elce di Sotto, 8 I-06123, Perugia, Italy}}\\ {$^{m}${RRC Kurchatov Institute, Kurchatov Sq.1,
123182 Moscow, Russia}}\\ {$^{n}${Dipartimento di Fisica Universit\`a and I.N.F.N., Genova, Via
Dodecaneso,33 I-16146 Genova, Italy}}\\ {$^{o}${KFKI-RMKI, Konkoly Thege ut 29-33 H-1121 Budapest,
Hungary}}\\ {$^{p}${Bell Laboratories, Lucent Technologies, Murray Hill, NJ 07974-2070, USA}}\\
{$^{q}${M.Smoluchowski Institute of Physics, Jagellonian University, PL-30059 Krakow, Poland}}}

%\thanks{
%{$^{*}$~Spokesman} \\ {$^{**}$~Project manager} \\ {$^{***}$~Operational manager} \\
%{$^{\circ}$~GLIMOS} \\ {$^{\diamond}$~Detector installation manager}\\  {$^{1}$~{Now at
%Massachusetts Institute of Technology, NW17-161, 175 Albany St. Cambridge, MA 02139} }\\
%{$^{2}$~Marie Curie fellowship at LNGS}\\ {$^{3}$~On leave of absence from St. Petersburg Nuclear
%Physics Inst. - Gatchina, Russia} \\ {$^{4}$~Now at Sudbury Neutrino Observatory, INCO Creighton
%Mine, P.O.Box 159 Lively, Ontario, Canada, P3Y 1M3} \\ {$^{5}$~Now at Institute for Nuclear
%Physics, Forschungszentrum Karlsruhe, Postfach 3640, 76021 Karlsruhe } \\ {$^{6}$~Now at Institute
%for Nuclear Research, Prospekt Nauki 47, MSP 03680, Kiev, Ukraine}\\ {$^{7}$~On leave of absence
%from KFKI-RMKI, Konkoly Thege ut 29-33 H-1121 Budapest, Hungary} \\ {$^{8}$~Present Address:
%Department of Physics, Virginia Polytechnic Institute and State University, Blacksburg VA  24061
%}\\ {$^{9}$~Center for Cosmological Physics, University of Chicago, 933 E. 56$^{th}$St., Chicago,
%IL 60637}\\ {$^{10}$~Deceased} }

\mail{derbin@mail.pnpi.spb.ru (A.Derbin)\\smirnov@lngs.infn.it (O.Smirnov)}

\abstract{ The Pauli exclusion principle (PEP) has been tested for nucleons ($n,p$) in $^{12}C$ and
$^{16}O$ nuclei, using the results of background measurements with the prototype of the Borexino
detector, the Counting Test Facility (CTF). The approach consisted of a search for $\gamma$, $n$,
$p$ and/or $\alpha$'s emitted in a non-Paulian transition of 1$P$- shell nucleons to the filled
1$S_{1/2}$ shell in nuclei. Similarly, the Pauli-forbidden $\beta^{\pm}$ decay processes were
searched for. Due to the extremely low background and the large mass (4.2 tons) of the CTF
detector, the following most stringent up-to-date experimental bounds on PEP violating transitions
of nucleons have been established: $\tau(^{12}C\rightarrow^{12}\widetilde{C}+\gamma) >
2.1\cdot10^{27}$ y, $\tau(^{12}C\rightarrow^{11}\widetilde{B}+ p) > 5.0\cdot10^{26}$ y,
$\tau(^{12}C(^{16}O)\rightarrow^{11}\widetilde{C}(^{15}\widetilde{O})+ n)
> 3.7 \cdot 10^{26}$ y, $\tau(^{12}C\rightarrow^{8}\widetilde{Be}+\alpha) >
6.1 \cdot 10^{23}$ y, $\tau(^{12}C\rightarrow^{12}\widetilde{N}+ e^- + \widetilde{\nu_e})> 7.6
\cdot 10^{27}$ y and $\tau(^{12}C\rightarrow^{12}\widetilde{B}+ e^+ + \nu_e)> 7.7 \cdot 10^{27}$ y,
all at 90\% C.L.
{}}

\titlerunning{New experimental limits on the violation....}
\authorrunning{Borexino coll., H.Back et al.}
\maketitle

\section{Introduction}
%%%%%%%%%%%%%%%%%%%%%%%%%%%%%%%%%%%%%%%%%%%%%%%%%%%%%%%%%%%%%%%%%%%%%%%%%%%%%%%%%%%%%%%%%%%%%%%
The exclusion principle was formulated by W.Pauli in 1925 and in its original form postulated that
only one electron with definite spin orientation can occupy each of the allowed Bohr orbits in an
atom. In this way PEP explains the regularities of the Periodic Table and atomic spectra. In modern
Quantum Field Theory (QFT), the PEP appears automatically for systems of identical fermions as a
result of the anti-commutativity of the fermion creation and annihilation operators.  Violation of
the PEP, as of the nonconservation of electric charge, would contradict modern quantum field
theory.

PEP has fundamental importance, but it was not extensively studied experimentally for 15 years
until the electron stability was tested. Goldhaber pointed out that the same experimental data
which were used to set a limit on the lifetime of the electron can be used to test the validity of
the PEP for atomic electrons \cite{Reines_Sobel}. Pioneering experiments were performed by Reines
and Sobel by searching for X-rays emitted in the transition of an L-shell electron to the filled
K-shell in an atom \cite{Reines_Sobel}, and by Logan and Ljubicic, who searched for $\gamma$-quanta
emitted in a PEP-forbidden transition of nucleons in nuclei \cite{Logan_Gamma}.

In 1980 Amado and Primakoff pointed out that in the framework of QFT, these PEP-violating
transitions are forbidden even if PEP-violation takes place \cite{Amado_Primakoff}. Later a
theoretical models describing a violation of PEP were constructed in
\cite{Ignatiev_Kuzmin}-\cite{Okun_JETP}, but it was found that even small PEP-violation leads to
negative probabilities for some processes \cite{Govorkov}. Critical studies of the possible
violation of PEP have been done both theoretically and experimentally by Okun
\cite{Okun_YFN},\cite{Okun_CNPP}.

One of the methods of testing PEP is the search for atoms or nuclei in a non-Paulian state; another
is the search for the prompt radiation accompanying non-Paulian transitions.

Violation of PEP in the nucleon system has been studied by searching for the non-Paulian
transitions with $\gamma$- \cite{Logan_Gamma},\cite{KAMIOKANDE},\cite{NEMO},
$p$-\cite{Ejiri},\cite{DAMA} and $n$-\cite{Kishimoto} emission, non-Paulian $\beta^+$, $\beta^-$-
decays \cite{Kekez},\cite{NEMO} and in nuclear $(p,p)$,$(p,\alpha)$-reactions on $^{12}$C
\cite{pp-reaction}.

The sensitivity of the forbidden transitions method is defined by the mass of the detector and by
the background level of detector. The extremely low background level and the large mass of the CTF
allowed setting new limits on the electron, neutrino and nucleon stability and neutrino
electromagnetic properties \cite{BOREX_EDecay} - \cite{BOREX_Heavy_Nu}. The approach used in the
search for nucleon and dinucleon disappearance \cite{BOREX_NNDecay} is close to the one used in the
present letter to search for PEP violation.

\section{Experimental set-up and measurements}
%%%%%%%%%%%%%%%%%%%%%%%%%%%%%%%%%%%%%%%%%%%%%%%%%%%%%%%%%%%%%%%%%%%%%%%%%%%%%%%%%%%%%%%%%%%%%%%
\subsection {Brief description of the CTF}
%%-------------------------------------------------------------------------------------
Borexino, a real-time 300~ton detector for low-energy neutrino spectroscopy, is nearing completion
in the Gran Sasso Underground Laboratory (see \cite{BORgen} and refs. therein). The main goal of
the detector is the measurement of the $^7$Be solar neutrino flux via $(\nu,e)$- scattering in an
ultra-pure liquid scintillator, but several other basic questions in astro- and particle physics
will also be addressed.

CTF, installed in the Gran Sasso underground laboratory, is a prototype of the Borexino detector.
Detailed reports on the CTF results were published elsewhere\cite{BORgen}-\cite{CTFlgt}, and only
the main characteristics of the set-up are outlined here.

The CTF consists of an external cylindrical water tank ($\oslash $11$\times $10 m; $\approx$1000 t
of water) serving as passive shielding for 4.2~m$^3$ of liquid scintillator (LS) contained in a
transparent nylon spherical vessel of $\oslash $2.0 m. High purity water with a radio-purity of
$\approx $$10^{-14}$ g/g~(U/Th), $\approx $$10^{-12}$~g/g~(K) and $<2\mu$Bq/{\it l} for
$^{222}$Rn is used for the shielding. The LS was purified to the level of $\simeq$10$^{-16}$~g/g in
U/Th contamination.

We analyze here the data of the second phase of the CTF detector (CTF2). The liquid scintillator
used at this stage was a phenylxylylethane (PXE, C$_{16}$H$_{18}$) with p-diphenylbenzene
(para-terphenyl) as a primary wavelength shifter at a concentration of 2~g/{\it l}, along with a
secondary wavelength shifter 1,4-bis-(2-methylstyrol)-benzene (bis-MSB) at 20~mg/{\it l}
\cite{PXE-paper}. The density of the scintillator is 0.99~kg/{\it l}. The scintillator principal
de-excitation time is less than 5~ns, which permits good position reconstruction. In the CTF2 an
additional nylon screen between the scintillator vessel and PMTs was installed, acting as a barrier
against
 penetration of external radon. The water volume of the CTF2 detector is instumented
with a \v{C}erenkov muon detector (muon veto system).
%\twocolumn %\vskip 0.5cm
\begin{figure}
%\begin{center}
\includegraphics[bb = 80 220 380 660, width=8cm,height=9cm]{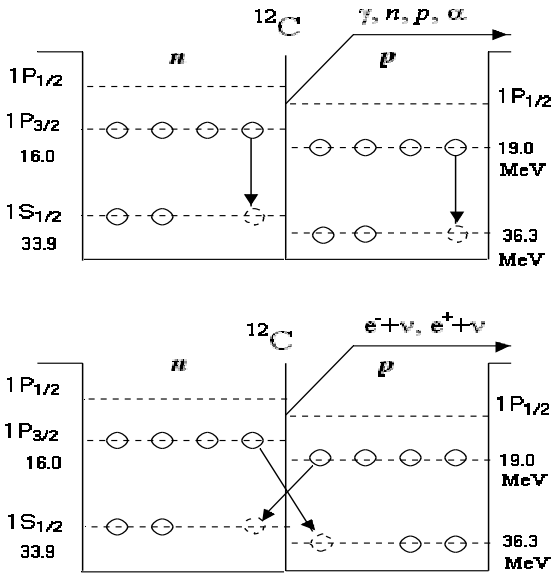}
%\mbox{\epsfig{figure=fig1.eps,width=14.5cm}}
\caption {Occupation of energy levels by protons and neutrons for the $^{12}$C ground state in a
simple shell model. Schemes of non-Paulian transitions of nucleons from the $P$-shell to the filled
$S$-shell: a) with $\gamma$-, $n$-, $p$- and $\alpha$-emission; b) with $\beta^+$-,
$\beta^-$-emission.}
%\end{center}
\end{figure}

The scintillation light is collected with 100~large phototubes (PMTs) fixed to a 7-~m diameter
support structure placed inside the water tank. The PMTs are fitted with light concentrators which
provide a total optical coverage of 21\%.

For each event the charge and time of every PMT hit are recorded. Each electronics channel is
supported by an auxiliary channel used to record events coming within a time window of 8.2~ms after
the trigger, which allows tagging of fast time-correlated events with a decrease of the overall
dead time of the detector. For longer delays, the computer clock is used, providing an accuracy of
$\approx$50~ms. Event parameters measured in the CTF2 include the total charge collected by the
PMTs during the 0--500~ns window, used to determine an event's energy; the charge in the 'tail' of
the pulse (48--548~ns) which is used to distinguish between $\alpha$ and $\beta$ events through the
pulse shape discrimination method; PMT timing, used to reconstruct the event's position; and the
time elapsed between sequential events, used to tag time-correlated events.

\subsection{Detector calibration}
%%----------------------------------------------------------------------------------------------
The energy of an event in the CTF detector is measured using the total collected charge from all
PMT's. In a simple approach, the response of the detector is assumed to be linear with respect to
the energy released in the scintillator. The coefficient $A$ linking the event energy and the total
collected charge is called the light yield (or photoelectron yield). Practically, the light yield
for electrons can be considered linear in energy only above $\sim$ 1~MeV. At low energies the
phenomenon of ``ionization quenching'' violates the linear dependence of the light yield versus
energy. The deviations from the linearity can be taken into account by the ionization deficit
function $ f(k_{B},E) $, where $k_B$ is the empirical Birks' constant \cite{Birks}. For the
calculation of the ionization quenching effect for the PXE scintillator, we used the KB program
from the CPC library \cite{Quenching}.

The ionization quenching effect leads to a shift in the position of the total absorption peak for
$\gamma$'s on the energy scale calibrated using electrons. In fact, the position of the 1461 keV
$^{40}K$ $\gamma$ in the CTF2 data corresponds to 1360 keV of energy deposited by an electron
\cite{BOREX_MMV}.

The energy calibration derived from the $^{14}$C
 $\beta$-spectrum gives
$A=3.72\pm$0.08~photoelectrons/(MeV $\times$ PMT) for high energy electrons depositing their energy
at the detector's center. \footnote{No measurements with high energy monoenergetic electrons are
available for the PXE scintillator. This is just convenient interpolation of the value obtained
fitting $^{14}$C spectrum, which allows to separate energy dependent part using ionization deficit
function $f(k_{B},E)$. The registered amount of light at any energy can be calculated as
$Y(E)=A\cdot f(k_{B},E)$. For 1~Mev electron $Y(1~$MeV)=3.54$\pm$0.08~p.e./(MeV~$\times$ PMT).}

The detector energy and spatial resolution were studied with radioactive sources placed at
different positions inside the active volume of the CTF2. A typical spatial 1$\sigma$ resolution is
10~cm at 1~MeV. The studies showed also that the total charge response of the detector can be
approximated by a Gaussian.  For energies $E\ge$1MeV (which are of interest here),
the relative resolution can be expressed as
$\sigma_E/E=\sqrt{3.8~keV / E +2.3\cdot10^{-3}}$ \cite{Smi00} for events uniformly distributed over
the detector's volume.

The energy dependence on the collected charge becomes non-linear for energies $E\geq5$~MeV because
of the saturation of the ADCs used. In this region we are using only the fact of whether or not
candidate events are observed, hence the mentioned nonlinearity doesn't influence the result of the
analysis.

More details on the energy and spatial resolutions of the CTF and ionization quenching for
electrons, $\gamma$ quanta and $\alpha$ particles were reported in
\cite{BOREX_EDecay},\cite{BOREX_MMV},\cite{Smi00}.

\subsection{Muon veto}
%%----------------------------------------------------------------------------------------------
The CTF2 was equipped with a water \v{C}herenkov muon veto system. It consists of 2 concentric
rings of 8~PMTs each, installed at the bottom of the tank. The radii of the rings are 2.4 and 4.8
~m.
Muon veto PMTs look upward and have no light concentrators. The muon veto system was optimized
in order to have a negligible probability of registering the scintillation events in the 250--800
~keV
$^{7}$Be neutrino energy region. The behaviour of the muon veto has been specially studied at
higher energies \cite{BOREX_NNDecay}. Experimental measurements with a radioactive source (chain of
$^{222}$Rn) \cite{CTFRn} gave the value $\eta(E)$=(1$\pm$0.2)\% in the 1.8--2.0~MeV region for the
probability $\eta(E)$ of identification of a scintillation event with energy $E$ in the LS as a
muon. The energy dependence of $\eta(E)$ was also calculated by a ray-tracing Monte Carlo method
accounting for specific features of the light propagation in the CTF2 which are detailed in
\cite{CTFlgt}. The calculated function was adjusted to reproduce correctly the experimental
measurements with the $^{222}$Rn source.

\section{Data analysis}
%%%%%%%%%%%%%%%%%%%%%%%%%%%%%%%%%%%%%%%%%%%%%%%%%%%%%%%%%%%%%%%%%%%%%%%%%%%%%%%%%%%%%%%%%%%%%%%
\subsection{Theoretical considerations}
%%---------------------------------------------------------------------------------------------
The non-Paulian transitions were searched for in nuclei of $^{12}$C contained in the scintillator
and $^{16}$O in the water shield of the  CTF2 detector, respectively. The nucleon level scheme of
$^{12}$C in a simple shell model is shown in Fig.1. The nucleon binding energies for the light
nuclei ($^{12}$C, $^{16}$O and others) were measured while studying $(p,2p)$ and $(p,np)$ proton
scattering reactions with 1 GeV energy \cite{PNPI}. The measured values for $1S_{1/2}$ and
$1P_{3/2}$ shells \cite{PNPI} together with $n$, $p$ and $\alpha$ separation energies \cite{Aud95}
are shown in table~1. For example, the measured values for the $1S_{1/2}$ shell of $^{12}$C are
$E_n(1S_{1/2},^{12}C)=36.3\pm0.6$~MeV and $E_p(1S_{1/2},^{12}C)=33.9\pm0.9$~MeV. These values
significantly differ from value $E_p(1S_{1/2},^{12}C)=39\pm1$ MeV extracted from
$(e,ep)$-scattering \cite{Francfurt}.

\begin{table}[!htbp]
\begin{center}
\caption{The separation energy $S_p$, $S_n$, $S_{\alpha}$ \cite{Aud95}, the nucleon binding energy
(with errors) of the 1$P_{3/2}$ and 1$S_{1/2}$ shells \cite{PNPI}, and the nuclear binding energy
$E_b$ (keV)\cite{Aud95}.}
\begin{tabular}{|l|c|c|c|c|}
   \hline
              & $^9$Be & $^{11}$B& $^{12}$C & $^{16}$O     \\ \hline
     $S_p$    & 16.9   &    11.2 & 16.0     &  12.3        \\
     $S_n$    & 1.66   &    10.7 & 18.7     &  15.7        \\
$S_\alpha$    &  2.5   &  8.7    & 7.4      &   7.2      \\ \hline 1$P_{3/2}(p)$ &  17.0~(0.2)  &
17.5~(0.5) & 16.0~(0.2)   &  18.0~(0.3)    \\ 1$P_{3/2}(n)$ &  18.1~(0.5)  & 18.4~(0.6)    &
19.0~(0.3) & 22.0~(0.4)
\\ \hline 1$S_{1/2}(p)$ &  27.7~(0.5)  & 33.5~(0.9)    &   33.9~(0.9)   &  39.8~(0.9)     \\ 1$S_{1/2}(n)$ &  29.2~(0.8)  &
34.5~(1.0) & 36.3~(0.6) &  42.2~(1.0)
\\ \hline
     $E_b$    &58164.9 & 76204.8 & 92161.8  & 127619.3      \\
   \hline
 \end{tabular}
\end{center}
\end{table}

The transition of a nucleon from the $P$-shell to the filled $S$-shell will result in excited
nuclei $^{12}\widetilde{C}$. The excitation energy corresponds to the difference of the binding
energies of nucleons on $S$- and $P$-shells. As one can see from the table, the energy release in
the non-Paulian transitions in $^{12}$C and $^{16}$O is comparable with separation energies $S_p$,
$S_n$, $S_{\alpha}$; hence, together with emission of $\gamma$-quanta, the emission of $n$, $p$ and
$\alpha$ is possible. Because of the uncertainties in the values of $E^{n,p}_{S_{1/2}}$, the
prediction of the branching ratio for the emission in each of the above mentioned channels has a
poor significance. For the case of the nucleon and dinucleon invisible decay in nuclei, the
branching ratio and spectra of the emitted particles were considered in \cite{Kamuskov}. In the
present paper we give the separate limits on the probabilities for each of the possible reactions.
The weak processes with a violation of the PEP (${\beta}^{+}$,${\beta}^{-}$-decays) \cite{NEMO},
\cite{Kekez} with a non-Paulian nucleon in the final state (on $1S_{1/2}$ shell) are considered as
well.

\subsection{Data selection}
%%---------------------------------------------------------------------------------------------
The candidate events, relevant for our studies, have to satisfy the following criteria: (1) the
event should occur in the volume of the detector and must not be accompanied by the muon veto tag;
the probability of detecting high energy events in LS has to be taken into account; (2) it should
be single (not followed by a time-correlated event) except in the case of neutron emission; (3) its
pulse shape must correspond to that of events caused by $\gamma$, $\beta$ or $\alpha$ particles
depending on the specific channel under study.

The experimental energy spectrum in CTF2, accumulated during 29.1~days of measurements (live time),
is shown in fig.~2. The trigger level was set at 21 fired PMT in a 30~ns window; the total count
rate at this threshold was 0.5~s$^{-1}$. The raw spectrum is presented on the top. The peak at
1.46~MeV, present in all spectra, is due to $\gamma$-quanta from $^{40}$K decays outside the
scintillator, mainly in the ropes supporting the nylon sphere. The peak-like structure at $\sim$6.2
~MeV is caused by saturation of the electronics by high-energy events.

\begin{figure}
\includegraphics[bb = 30 90 500 760, width=8cm,height=10cm]{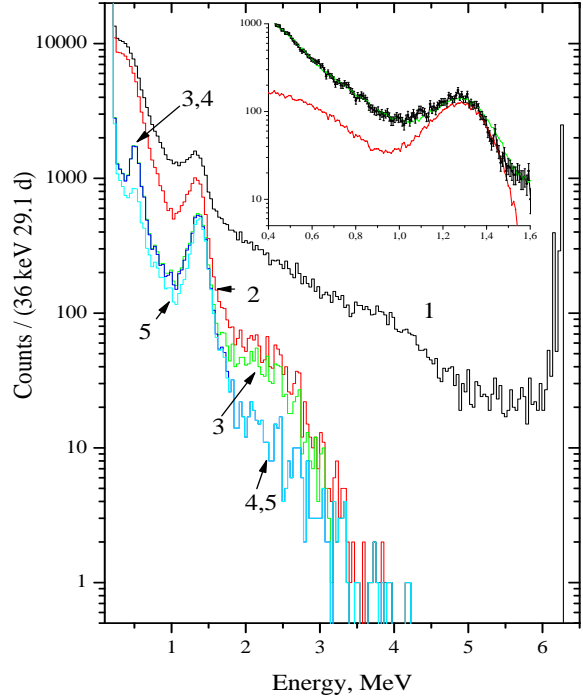}
\caption {
 Background energy spectra of the 4.2 ton Borexino CTF2 detector measured over 29.1
days. From top to bottom: (1) the raw spectrum; (2) with muon veto cut; (3) non-muon events inside
the radius $R\leq$100 cm; (4) pairs of correlated events (with time interval $\Delta t\leq$8.2 ms
between signals) are removed; (5) $\alpha/\beta$ discrimination is applied to eliminate any
contribution from $\alpha$ particles.  In the inset, the simulated response function for external
$^{40}$K $\gamma$'s is shown together with the experimental data.}
\end{figure}

The second spectrum is obtained by applying the muon cut, which suppressed the background rate by
up to two orders of magnitude, depending on the energy region. No events with energy higher than
4.5~MeV passed this cut. In the next stage of the data selection
we applied a cut on the reconstructed radius. We used a $R\leq$100~cm (radius of the inner vessel)
cut aiming to reduce significantly the surface background events (mainly due to the $^{40}$K decays
outside the inner vessel) and leave the events uniformly distributed over the detector volume. The
efficiency of the cut has been studied with MC simulation and lies in the range of
$\epsilon_R=0.76-0.80$ in the energy region 1--2~MeV. The time-correlated events (occurring in the
time window $\Delta t\leq$~8.2~ms) were also removed (spectrum~4). Suppression of non-correlated
events is negligible (0.4\%) due to low count rate. Additional $\alpha /\beta$ discrimination
\cite{CTFgen} was applied to eliminate contribution from $\alpha$ particles (spectrum~5 in fig.~2).
The  loss of $\beta$-particles for $\alpha$ identification efficiency 85\% is less than 2\% for the
1~MeV energy region \cite{PXE-paper}.

The selection and treatment of data (spatial cuts, analysis of an event's pulse shape to
distinguish between electrons and $\alpha$ particles, suppression of external background by the
muon veto system, etc.) is similar to that in ref. \cite{BOREX_EDecay},\cite{BOREX_MMV},
\cite{BOREX_NNDecay},\cite{BOREX_Heavy_Nu}.

\subsection{Simulation of the response functions}
%---------------------------------------------------------------------------------------------
Due to the complex phenomena of light propagation in a large volume scintillator detector, the
precise modeling of the detector response is a complicated task. Among the problems, it is worth
mentioning the wavelength dependence of the processes involved in light propagation;
reflection/refraction on the scintillator/water interface; the light reflection on the
concentrators; etc. \cite{CTFlgt}. The non-spherical shape of the inner vessel, deformed by the
supporting strings, is an additional source of uncertainty. The need to follow each of the 12000
photons emitted per 1 MeV electron event makes tracing MC code very slow.

We developed a fast reliable code, based on the measurements with the detector. The perfect sets of
data for the code tuning are the $^{14}$C $ \beta - $decay data and the easily identified $
\alpha$'s from the radon decay in the scintillator volume. The code has two parts: the
electron-gamma shower simulation (EG code) and the simulation of the registered charge and position
(REG code). The EG code generates a random- position event with a random initial direction (for
$\gamma$'s) and follows the gamma- electron shower using the EGS-4 code\cite{EGS4}. The low-energy
$e$ and $\alpha$ are not propagated in the program and are considered to be point-like, with the
position at the initial coordinates.

The mean registered charge corresponding to the electron's energy $ E_{e} $ is calculated by
\begin{equation}
\label{DeltaQ} Q_{e}=A\cdot E_{e}\cdot f(k_{B},E_{e})\cdot f_{R}(r),
\end{equation}
where $ f_{R}(r) $ is a radial factor taking into account the dependence of the registered charge
on the distance from the detector's center, and $ f(k_{B},E_{e}) $ is the quenching factor for
electrons. The method used to obtain $ f_{R}(r) $ is described in \cite{fR}.

From the analysis of the PXE data, the quenching factor $ k_{B}=(1.5\pm0.1)\cdot10^{-3} $ was found
to satisfy experimental data \cite{BOREX_MMV}. The value is in agreement with the high statistics
fit of the $^{14}$C $ \beta - $spectrum. The presence of the strong $\gamma$ line of 1.46~MeV in
the CTF2 data was used to check the method: first, the quenching factor was extracted from the
$^{14}$C $ \beta - $spectrum, and then the $^{40}$K $\gamma$'s were simulated. The position of
the peak in the model agreed with the real data to within 1\% accuracy.

The mean registered charge corresponding to the $\alpha$ of energy $ E_{\alpha} $ is calculated by
\begin{equation}
\label{DeltaQ1} Q_{\alpha }=A\cdot E_{\alpha }\cdot f_{\alpha }(E_{\alpha})\cdot f_{R}(r),
\end{equation}
where $ f_{\alpha }(E_{\alpha}) $ is the quenching factor for $\alpha$'s. The following
approximation of the quenching factor$ f_{\alpha }(E_{\alpha}) $ was found on the basis of
laboratory measurements for a scintillator based on pseudocumene (PC):
\begin{equation}
f_{\alpha }^{PC}(E_{\alpha})=\frac{1}{a-b\cdot E_{\alpha}}.
\end{equation}
with $a$~=~20.4 and $b$~=~1.3. The measurements of the $ f_{\alpha }(E_{\alpha}) $ for PXE were
performed in laboratory \cite{Neff} and analyzing delayed spectra of the CTF2 \cite{PXE-paper}. It
was found that an $ \alpha $-particle with energy 7.69~MeV is quenched to an equivalent $ \beta
$-energy of $ 950\pm12$~keV. Other reference points were found using the peaks corresponding to 3
$ \alpha $-particle of 5.3, 5.49 and 6.02~MeV correspondingly. Using the same form of approximation
as for PC, we found $a$~=~16.2 and $b$~=~1.1 for the PXE scintillator.

The $\gamma$'s were propagated using EGS-4 code \cite{EGS4}. As soon as an electron of energy $
E_{e} $ is to appear inside the scintillator, the corresponding charge is calculated:
\begin{equation}
\label{DeltaQ2} \Delta Q_{i}=A\cdot E_{e_{i}}\cdot f(k_{B},E_{e_{i}})\cdot f_{R}(r_{i});
\end{equation}
total mean collected charge is defined when the $\gamma$ is discarded by the EG code, summing
individual deposits:
\begin{equation}
Q_{\gamma }=\sum \Delta Q_{i}.
\end{equation}
The weighted position is assigned to the final $\gamma$:
\begin{equation}
\label{Formula:xw} x_{w}=\frac{\sum \Delta Q_{i}\cdot x_{i}}{\sum
\Delta Q_{i}},
\end{equation}
where $ \Delta Q_{i} $ and $ x_{i} $ are the charge deposited for the $ i^{th} $ electron at the
position ($x_{i},y_{i},z_{i}$). The analogous rule is applied for the $ y_{w} $ and $ z_{w} $
coordinates.

In the next step a random charge is generated in accordance to the normal distribution with a mean
value of $ Q=\sum \Delta Q $ and with variance $ \sigma _{Q}=\sqrt{(1+\overline{v_{1}})\cdot Q} $.
The instrumental parameter $\overline{v_1}$ is the relative variance of the single photoelectron
charge response averaged over all PMTs of the detector. It was defined independently from the
measurements with radon source inserted in the detector and from averaging the relative variances
of the single photoelectron response obtained during the acceptance tests \cite{Smi00}. Both method
give $\overline{v_1}=0.34\pm0.01$.

Finally, the radial reconstruction is simulated taking into account energy dependence of the
reconstruction precision. It is assumed that the reconstruction precision is defined by the number
of PMTs fired in an event and that reconstruction precision doesn't depend on the position. These
two facts are in agreement with the measurements using the artificial radon source inserted in the
CTF1 and CTF2 detectors \cite{CTFgen}.

The fit of the radial distribution of the $\alpha$-particle events with energy $E_{\alpha}$=7.69~MeV
(equivalent electron energy $E_e$=950~keV) gives $\sigma _{R}=13.8$~cm. If we assume that the
reconstruction precision is defined by the mean number $N$ of fired channels, then the
$\sigma_R(E)$ for an event of energy $E$ is:
\begin{equation}
\label{Formula:RecPrecision} \sigma _{R}(E)=\sigma _R(950~keV)\cdot
\sqrt{\frac{N(950~keV)}{<N(E)>}}.
\end{equation}
where mean number of fired PMTs for $E$=950 keV is $N(950~keV)$~=~91 (of the total 100).
The number of the fired
 channels $N(E)$ was simulated for every event assuming a Poisson
distribution of photoelectrons
 registered on each PMT.

\section{Results}
%%%%%%%%%%%%%%%%%%%%%%%%%%%%%%%%%%%%%%%%%%%%%%%%%%%%%%%%%%%%%%%%%%%%%%%%%%%%%%%%%%%%%%%%%%%%%%%
\subsection{Limits on non-Paulian transitions in $^{12}$C      %%%%% GAMMA
and $^{16}$O with emission of $ \gamma$. }                     %%%%% GAMMA
%%----------------------------------------------------------------------------------------------

As follows from table~1, the energy difference for the nucleon transition from the shell $1P_{3/2}$
to the shell $1S_{1/2}$ is $\simeq$~17.5~MeV for $^{12}$C. The response functions of the CTF2
to the $\gamma$ of this energy were simulated by  the MC methods described in the previous
subsection. The energy difference for the same transition in the case of  $^{16}$O corresponds to
$\simeq$21~MeV. The uniformly distributed $\gamma$'s of this energy were simulated in the 1~m-
thick layer of water surrounding the scintillator. Both response functions are shown in Fig.~3
before and after the muon veto suppression.

\begin{figure}
\includegraphics[bb = 30 90 500 760, width=8cm,height=10cm]{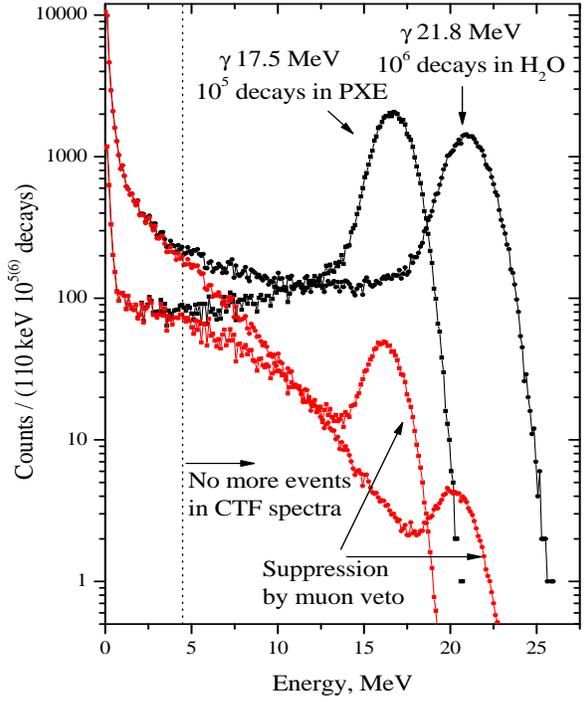}
\caption {The expected response functions of the detector for
$^{12}C\rightarrow^{12}\widetilde{C}+\gamma$ decays in the liquid  scintillator and
$^{16}O\rightarrow^{16}\widetilde{O}+\gamma$ decays in the water shield before and after muon veto
suppression are shown.}
\end{figure}

The limit on the probability of transitions $^{12}$C$\rightarrow^{12}\widetilde{C}+\gamma$ and
$^{16}$O$\rightarrow ^{16}\widetilde{O}+\gamma $ violating the PEP are based on the experimental fact
of observing no events with energy higher than 4.5~MeV passing muon veto cut. The
lower limit on PEP violating transitions of nucleons from $P$-shell to the occupied
$1S_{1/2}$-shell was obtained using the formula
\begin{equation}
\label{TLimit} \tau \geq \varepsilon _{\Delta E}\frac{N_{N}N_{n}}{S_{lim}}T,
\end{equation}
where $ \varepsilon _{\Delta E} $ is the efficiency of registering an event in the energy interval
$ \Delta E $, $ N_{N}$ is the number of nuclei under consideration, $N_{n}$ is the number of
nucleons ($n$ and/or $p$) in the nuclei for which the non-Paulian transitions are possible, $ T $
is the total time of measurements, and $ S_{lim} $ is the upper limit on the number of candidate
events registered in the $ \Delta E $ energy interval and corresponding to the chosen confidence
level.

The efficiency of 17.5~MeV $ \gamma  $ detection $ \varepsilon _{\Delta E}=4.3\cdot 10^{-2}$ was
determined in a MC simulation, taking into account the suppression of the high energy events by the
muon veto system (fig.~3). The number of $ ^{12}C $ target nuclei in $ 4.17 $ tonnes of liquid
scintillator based on PXE is $ N_{N}=1.89\cdot 10^{29} $ (taking into account the isotopic
abundance of $^{12}$C). The number of nucleons on the $ P $-shell is $ N_{n}=8 $, the total data
taking time is $T=0.080$~y, and the upper limit on the number of candidate events is
$S_{lim}=2.44$ with 90\%~C.L. in accordance with the Feldman-Cousins procedure \cite{Feldman},
recommended
 by the Particle Data Group.

The $^{16}$O nucleus has 8 nucleons on its $1P_{3/2}$ and 4 nucleons on its $1P_{1/2}$ shell. The
values of the binding energy of the nucleons on $1P_{1/2}$-shell are $E_p(1P_{1/2})=13.4\pm0.4$~MeV
and $E_{n}(1P_{1/2})=16.2\pm0.3$~MeV \cite{PNPI}. The values of $\varepsilon_{\Delta E}$ were
calculated for $E_{\gamma}=21.8$~MeV, 26~MeV
($1P_{3/2}\rightarrow1S_{1/2}$,$1P_{1/2}\rightarrow1S_{1/2}$ transitions) and for the two
$\gamma$-quanta in the cascade $E_{\gamma}=21+5$~MeV. These values are $5.7\cdot10^{-3}$,
$5.4\cdot10^{-3}$, and $2.2\cdot10^{-2}$ correspondingly. Because of the unknown branching ratio,
the worst possible total efficiency $\varepsilon_{\Delta E}=5.6\cdot10^{-3}$ was adopted. The
number of target $^{16}$O nuclei in the 1~meter thick layer of water shielding is
$N_{N}=9.8\cdot 10^{29}$ and the upper limit on the number of the candidate events corresponding to
the 90\% c.l. is the same as in the previous case, $S_{lim}=2.44$. The limits obtained using the
cited numbers are:
\begin{equation}
\tau (^{12}C\rightarrow ^{12}\widetilde{C}+\gamma )\geq 2.1\cdot10^{27}\; y,
\end{equation}
\begin{equation}
\tau (^{16}O\rightarrow ^{16}\widetilde{O}+\gamma )\geq 2.1\cdot 10^{27}\; y,
\end{equation}
with a combined limit of $ \tau \geq 4.2\cdot 10^{27}\; y $ for the 90\% c.l. This result is
stronger than the ones obtained with the NEMO-2 detector  $\tau (^{12}C\rightarrow
^{12}\widetilde{C}+\gamma )\geq 4.2\cdot 10^{24}\; y$ \cite{NEMO},
 and the Kamiokande detector $\tau (^{16}O\rightarrow
^{16}\widetilde{O}+\gamma )\geq 2.7\cdot 10^{27}\; y$ \cite{KAMIOKANDE}.

\subsection{Limits on non-Paulian transitions in $^{12}$C       %%%% PROTON
with proton emission $^{12}$C$\rightarrow ^{11}\widetilde{B}+p$.}          %%%% PROTON
%%---------------------------------------------------------------------------------------------
Energy released in these transitions is the difference between the binding energies of the final
and initial nuclei:
\begin{center}
$Q(^{12}C\rightarrow ^{11}\widetilde{B}+p)=M(^{12}C)-M(^{11}\widetilde{B})-m_{p}=$
\end{center}
\begin{equation}
-E_{b}(^{12}C)+E_{b}(^{11}\widetilde{B});
\end{equation}
with the evident notations. The binding energy of the non-Paulian nuclei with 3 neutrons $
E_{b}(^{11}\widetilde{B}_{n}) $ or 3 protons $ E_{b}(^{11}\widetilde{B}_{p}) $ on the S-shell can
be evaluated considering the binding energy $ E_{b}(^{11}B) $ and the difference between the
binding energies of nucleons on the S-shell $ E_{n,p}(S_{1/2}) $ and the binding energy of the last
nucleon $ S_{n,p}(^{11}B) $:
\begin{equation}
E_{b}(^{11}\widetilde{B}_{n,p})\simeq E_{b}(^{11}B)+E_{n,p}(1S_{1/2})-S_{n,p}(^{11}B).
\end{equation}
Using the data of Table 1, one can obtain $Q(^{12}$C$\rightarrow^{11}\widetilde{B_{p}}+p)=6.3 $ MeV
and $Q(^{12}$C$\rightarrow ^{11}\widetilde{B_{n}}+p)=7.8$~MeV. Taking into account the recoil
energy of the nuclei and experimental errors of $ E_{n,p}(S_{1/2}) $ from table~2, the energy of
the proton released in these non-Paulian transitions is $ E_{p}=5.8(7.2)\pm1.0$~MeV.

The light yield for the protons in the range of 0.6-6.0~MeV was measured for the NE213 scintillator
using recoil protons from the $(n,p)$ elastic scattering \cite{p-quenching}. The measured data were
approximated by the formula relating energy release of protons $E_{p}$ and electrons $E_{e}$:
\begin{equation}
E_{e}=0.034\cdot E_{p}^{2}+ 0.311\cdot E_{p}- 0.109
\end{equation}

The light yield for a proton with energy $E_{p}$=5.8(7.2)~MeV corresponds to an electron energy of
$E_{e}$=2.8(3.9)$\pm$ 0.5~MeV. It means that the proton peak can be found in the energy interval
2.0-4.7~MeV with 90\% probability. The uncertainty of the peak position is few times higher than
energy resolution of CTF2 ($ \sigma_E $=130 keV for $E_e$ = 2~MeV) and covers errors due to using
NE213 data instead of that for PXE.

To establish limits on the probability of these non-Paulian transitions in $^{12}$C, we use the
formula (\ref{TLimit}). Because of uncertainty in the $p$ peak position, $S_{lim}$ was determined
using a very conservative approach: it was defined as the number of events $N$ inside the
2$\sigma_E$ window ($\varepsilon_{\triangle E}$=0.68) which can be excluded at a given confidence
level ($N$+1.28$\sqrt{N}$ for 90\%~c.l.). This procedure was used for the wide energy interval
2.0--4.7~MeV. The maximum value of $S_{lim}=130$ at 90\%~c.l. (and the least stringent limit on
life-time) was obtained for the energy interval 2.0--2.26~MeV. The lower limit on the life-time was
found from formula (\ref{TLimit}) taking into account the efficiency of radial cut
$\epsilon_{R}$=0.8:
\begin{equation}
\tau (^{12}C\rightarrow ^{11}\widetilde{B}+p)\geq 5.0\cdot 10^{26}\; y\: (90\%\:
c.l.).\end{equation}
This result is stronger than ones obtained with the 300~kg NaI ELEGANT~V
detector $ \tau (^{23}Na,^{127}I\rightarrow ^{22}\widetilde{Ne},^{126}\widetilde{Te}+p)\geq
1.7\cdot 10^{25}\; y\: (90\%\: c.l.) $ for protons with $E_p$$\geq$18~MeV, and with the 100 kg NaI
DAMA detector $ \tau (^{23}Na, ^{127}I\rightarrow ^{22}\widetilde{Ne},^{127}\widetilde{Te}+p)\geq
(7-9)\cdot 10^{24}\; y\: (90\%\: c.l.) $ for protons with $E_p$$\geq$10~MeV.

\subsection{Limits on non-Paulian transitions in $^{12}$C with               %%%%%% ALPHA
$\alpha$-particle emission $^{12}$C$\rightarrow ^{8}\widetilde{Be}+\alpha$.} %%%%%% ALPHA
%%--------------------------------------------------------------------------------------------
The binding energy of an $ \alpha  $-particle in $^{12}$C nuclei is as low as 7.4~MeV. The energy
released in the transition is the difference between the binding energies of the final and initial
nuclei:
\begin{equation}
Q(^{12}C\rightarrow ^{8}\widetilde{Be}+\alpha
)=-E_{b}(^{12}C)+E_{b}(^{8}\widetilde{Be})+E_{b}(^{4}He).
\end{equation}
The binding energy of nucleons on the S-shell of $^{8}\widetilde{Be} $ can be obtained using
experimental values $E_{n,p}(1S_{1/2})$ for the isotope $^{9}$Be. The binding energies for the
non-Paulian nuclei $^{8}\widetilde{Be_n} $ and $^{8}\widetilde{Be_p} $ calculated thus give values
$ Q\simeq2.9\pm0.9$~MeV and $ Q\simeq3.0\pm0.6 $, respectively. As the result, the
$\alpha$-particles from the decay can be found in the energy interval 1.0 - 3.0~MeV with 90\%
probability. In accordance with (3), the light yield for an $\alpha$ with energy between 1.0--3.0
MeV corresponds to an electron in the energy range 70 - 230~keV. The CTF2 efficiency of
$\alpha/\beta$ discrimination was not studied in this energy region and $\alpha/\beta$ selection
was not used for reaction (15). The dominant part of the background in this range is the
$\beta$-activity of $^{14}$C. For the energy window (230 keV$\pm\sigma_E$, $\sigma_E$=30~keV) the
value $S_{lim}$ is 3400 at 90\%~c.l (spectrum~4 on fig.~2) . Taking into account the efficiency of
the radial cut $\epsilon_R$=0.67, the lower limit on lifetime for decay~(15) is
$\tau\geq$1.6$\cdot$10$^{25}$ y. Our results on electron stability, obtained on the same
experimental data, can be used to set stronger limit on the peak near the endpoint of the
$\beta$-spectrum \cite{BOREX_EDecay}.

For an $\alpha$-particle with $E_{\alpha}$~=~1~MeV ($E_e$=70~keV) the limit is weaker. Measurements
with low threshold (6 fired PMTs, or $\approx$ 20 keV) were performed with 3 tons of PXE. The dead
time of the system  with this low threshold was 43\%. At the energy 70~keV, the number of counts in
the $\beta$-spectrum in the interval 2$\sigma_E$ is (6.6$\pm$0.2)$\cdot$10$^{4}$ d$^{-1}$, where
error of about 3\% includes both systematic and statistic effects \cite{PXE-paper}. For values
$S_{lim}$=3.3$\cdot$10$^{3}$, $\varepsilon_{\triangle E}$=0.68, $N_N$=1.35$\cdot$10$^{29}$ and $T$=
2.74$\cdot$10$^{-3}$ y, one can obtain
\begin{equation}
\tau (^{12}C\rightarrow ^{8}\widetilde{Be}+\alpha )\geq 6.1\cdot 10^{23}\: y\: (90\%\: c.l.).
\end{equation}

\subsection{Limits on non-Paulian transitions in $^{12}$C and $ ^{16}O$ with neutron emission:
$ ^{12}C\rightarrow^{11}\widetilde{C}+n$ , $ ^{16}O\rightarrow ^{15}\widetilde{O}+n$.}     %%%NEUTRON
%%-----------------------------------------------------------------------------------------%%%NEUTRON
The energy released in the decay $ ^{12}$C$\rightarrow^{11}\widetilde{C}+n $ is equal to the
difference of the binding energies of $^{12}$C and $^{11}\widetilde{C} $:
\begin{equation}
Q(^{12}C\rightarrow ^{11}\widetilde{C}+n)=-E_{b}(^{12}C)+E_{b}(^{11}\widetilde{C}).
\end{equation}
The binding energy of non-Paulian nuclei $ ^{11}\widetilde{C} $ can be defined as $
E_{b}(^{11}\widetilde{C})\simeq E_{b}(^{11}C)+E_{n,p}(S_{1/2},^{11}C)-S_{n,p}(^{11}C) $. The
nucleus $^{11}$C is unstable. Its nucleon separation energies are $ S_n$=13.1 and $ S_p$ = 8.7
MeV. If one assumes that the binding energies of the nucleons of $ ^{11}C $ on the $ S_{1/2} $
shell are close to those of $^{12}$C nuclei, then the energies released in the process~(17) are
$Q=6.5$ and 4.5~MeV for the nuclei $^{11}\widetilde{C_{p}} $ and $ ^{11}\widetilde{C_{n}} $ in
their final states, respectively. For decays in the water, the mean neutron energy defined in the
analogous way is $\simeq18$~MeV.

The resulting neutrons are thermalized in hydrogen-rich media (organic scintillator or water). The
lifetime of neutrons in water and/or scintillator is the order of some hundreds of $\mu$s, after
which they are captured by protons. The cross section for capture on a proton for a thermal neutron
is $0.3$~barns. The cross sections are much smaller for capture on the $ ^{16}O $ and $ ^{12}C $
nuclei: $ \sigma _{c}(^{12}$C=3.5~mbarns and $ \sigma _{c}(^{16}$O=0.2~mbarns. Capture of
thermal neutrons $ n+p\rightarrow d+\gamma  $ is followed by $ \gamma  $- emission with 2.2~MeV
energy. The background levels measured in CTF2 at this energy have been used to obtain an upper
limit on the number of $ \gamma $'s with 2.2~MeV energy, and as a result, a limit on the
probability of neutron production in the reactions $^{12}$C$\rightarrow ^{11}\widetilde{C}+n $ ,
$^{16}$O$\rightarrow ^{15}\widetilde{O}+n $. Protons scattered during the thermalization with energies
of some~MeV can be registered by the detector, hence the sequential events were not cut out in the
data selection. As a result, the main contribution to the background in the 2~MeV region was
attributed to the decays $^{214}$Bi$\rightarrow^{214}$Po that were not suppressed by the delayed
coincidence cut (see Fig.~4).

\begin{figure}
\includegraphics[bb = 30 30 500 760, width=8cm,height=11cm]{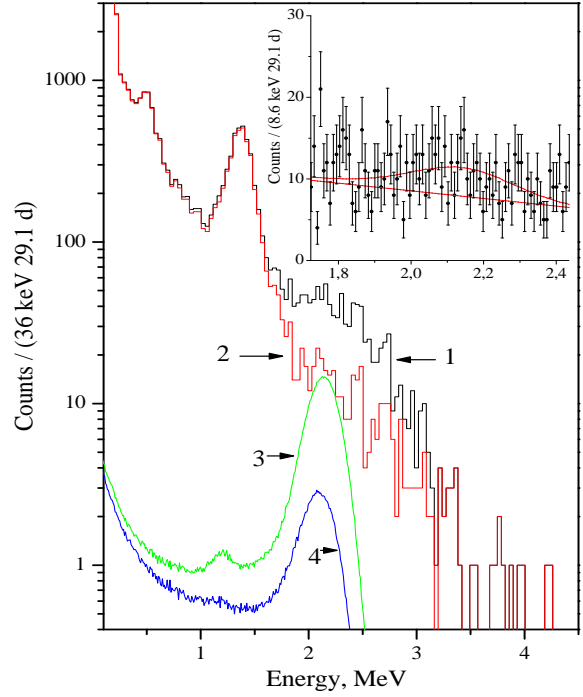}
\caption {Background energy spectra of the 4.2 ton Borexino CTF2 detector measured over 29.1 days:
(1) with muon veto, radial cut ($R$$\leq$100 cm) and $\alpha/\beta$ discrimination applied; (2)
pairs of correlated events (with time interval $\Delta t$$\leq$8.2 ms between signals) are removed;
(3) the expected response functions of the detector for $\gamma$'s with energy 2.2 MeV due to
$(n,p)$-capture in the scintillator and (4) in the water. Corresponding mean lifetimes for the
 $^{12}C\rightarrow^{11}\widetilde{C}+ n$ and  $^{16}O\rightarrow^{15}\widetilde{O}+ n$ decays
are $\tau_{lim}=3.7\cdot 10^{26}$ y. In the inset the fitting function in the energy interval
1.7--2.5 MeV is shown.}
\end{figure}
The analysis was made under the assumption that the mean lifetime of the nucleons is the same in
$^{12}$C and $^{16}$O nuclei, and that $n$-emission decays within both of them contribute to the
experimental spectrum simultaneously. The density of the PXE scintillator practically matches that
of water. The density of nucleons that can cause the non-Paulian transitions with neutron emission
is $ 2.7\times 10^{23}$~cm$^{-3}$ for PXE and $4.0\times 10^{23}$~cm$^{-3}$ for water. The
response function for $ 10^{6} $ initial $\gamma$'s generated in the liquid scintillator volume,
and for $10^{7}$ $\gamma$'s generated in the water layer of 100~cm, are shown in Fig.~4. The ratio
of the volumes of PXE and water where the $ \gamma $-events were simulated is $1:7$; the
difference in densities of hydrogen atoms capturing neutrons is small ($ 5.1\times 10^{22}$ cm$^{-3}$
for PXE and $ 6.7\times 10^{22}$~cm$^{-3}$ for water); hence the response function in
Fig.~4 corresponds to practically equal probabilities for the non-Paulian transitions for nucleons
in $^{12}$C and $^{16}$O nuclei. The response function of the CTF2 to the $\gamma$'s of 2.2~MeV
energy was obtained using the MC model. The shift in the positions of the peaks from 2.2~MeV toward
lower energies is a result of the ionization quenching of Compton electrons with a low energy.

The background in the 1.7-2.5~MeV region is a linear function of energy (Fig.~4). Because the peak
position is well known, the maximum likelihood method was used to define the intensity of the peak
corresponding to the 2.2~MeV $ \gamma  $. Experimental background was modeled as a linear function
plus the additional contributions of the response functions for 2.2~MeV $\gamma$ originating from
the PXE and water. The results of fitting are shown in the inset to Fig.~4. The minimum value of
$\chi^{2}=83.9/90$ corresponds to the total of 100 $\gamma$ events inside LS. For the 90\%
confidence level, the corresponding limiting number of $\gamma$ originating from the scintillator
is 260. Taking into account the efficiency of the radial cut $ \epsilon _{R}=0.80 $, the total
number of captured neutrons in the CTF2 can be limited by $ N\leq$2.7$n$/(d$\cdot$t). This value is
close to the expected rate of neutrons production by muons $N\approx$1$n$/(d$\cdot$t)
\cite{Ryazhskaya}.

Finally, the limit on PEP violating transitions of nucleons in $^{12}$C and $^{16}$O nuclei with
neutron emission is
\begin{equation}
\tau (^{12}C(^{16}O)\rightarrow ^{11}\widetilde{C}(^{15}\widetilde{O})+n)\geq 3.7\times 10^{26}\:
y\: (90\%\: c.l.).
\end{equation}
This result is 6 orders of magnitude stronger than the one obtained through searching for neutron
emission in Pb : $ \tau($Pb$\rightarrow \widetilde{Pb}+n)\geq 1.0\cdot 10^{20}\; y\: (68\%\:
c.l.) $ \cite{Kishimoto}.

\subsection{Limits on non-Paulian $ \beta ^{-}$transitions                    %%%%%BETA-
in $^{12}$C: $ ^{12}$C$\rightarrow ^{12}\widetilde{N}+e^{-}+\overline{\nu} $.}            %%%%%BETA-
%%------------------------------------------------------------------------------------------------
The nucleus $^{12}$N is unstable; it decays  via $^{12}$N$\rightarrow ^{12}C+e^{+}+\nu $ with an
energy release of $ Q=17.3 $~MeV. The inverse process $^{12}$C$\rightarrow
^{12}\widetilde{N}+e^{-}+\overline{\nu}  $ is possible if the binding energy of the non-Paulian
nucleus $ E_{b}(^{12}\widetilde{N}) $ is increased in comparison to the binding energy of the
normal $^{12}$N nucleus by a value exceeding $ Q $. The energy released in the reaction
$^{12}$C$\rightarrow ^{12}\widetilde{N}+e^{-}+\overline{\nu} $ is
\begin{equation}
Q=m_{n}-m_{p}-m_{e}-E_{b}(^{12}C)+E_{b}(^{12}\widetilde{N}).
\end{equation}
The value of $ E_{b}(^{12}\widetilde{N}) $ can be approximated by $E_{b}(^{12}\widetilde{N})\simeq
E_{b}(^{12}N)+E_{p}(S_{1/2},^{12}N)-S_{p}(^{12}N). $ The separation energy of the proton in
$^{12}$N has a very low value, $ S_{p}(^{12}N)=0.6 $ MeV. The value of $ E_{p}(S_{1/2},^{12}N) $
can be approximated by the mean value of the binding energies on the $ S_{1/2} $ shell for two
neighboring nuclei: $ E_{p}(S_{1/2},^{12}$N$)\simeq 0.5\cdot
(E_{p}(S_{1/2},^{12}$C$)+E_{p}(S_{1/2},^{16}$O$))=36.8$~MeV. Hence, the expected value of $ Q $ is
$18.9$~MeV.

The shape of the $ \beta ^{-} $ spectrum with end-point energy 18.9~MeV is shown in Fig.~5. The
limit on the probability of this transition was based on the fact of observing no events with
$E_e\geq$4.5~MeV not accompanied by a muon veto signal. As noted above, it is necessary to take
into account the probability $\eta(E_e)$ of the muon veto triggering for the high energy events in
scintillator. The obtained efficiency of detection of electrons with energies $E_{e}>4.5$~MeV is
$ \Delta \varepsilon =0.31 $. The limit on the lifetime of neutrons ($N_n$=4) in $^{12}$C with
respect to the transitions violating the PEP is
\begin{equation}
\tau (^{12}C\rightarrow ^{12}\widetilde{N}+e^{-}+\overline{\nu} )\geq 7.6\cdot 10^{27}\: y\:
(90\%\: c.l.).
\end{equation}
This result is 3 orders of magnitude stronger than the one obtained by NEMO-2,
$\tau(^{12}$C$\rightarrow ^{12}\widetilde{N}+e^{-}+\overline{\nu} )\geq 3.1\cdot 10^{24}\: y\: (90\%\:
c.l.) $ \cite{NEMO}.

\begin{figure}
\includegraphics[bb = 30 30 500 760, width=8cm,height=10cm]{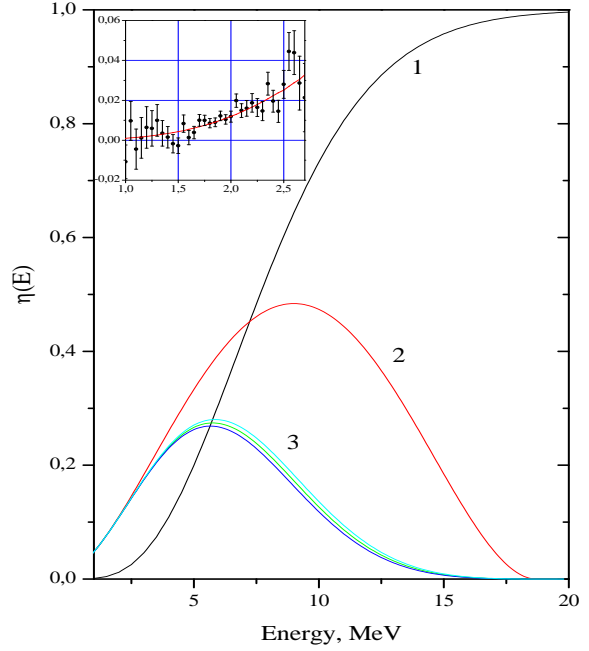}
\caption {Probability of identification of an event with energy $E$ in the scintillator by the muon
veto $\eta(E)$ (1). The $\beta$ spectra of $^{12}C\rightarrow^{12}\widetilde{N}+
e^{-}+\overline{\nu_{e}}$ without (2) and with (3) suppression by the muon veto are also shown in
arbitrary units.  In the inset, the $\eta(E)$ function together with the experimental data taken
with the radon source are presented. The tagged events of $^{214}$Bi--$^{214}$Po at the energy
$E$=1.9 MeV are 'seen' by the muon veto system with an efficiency of $\eta(1.9\:MeV)=0.01$.}
\end{figure}

The data available from the LSD detector \cite{LSD} situated in the tunnel under Mont Blanc allows
obtaining a qualitative limit for this decay comparable to ours. In \cite{Kekez}, it is claimed
that only 2~events were observed with energies higher than 12~MeV during 75~days of data taking
with the detector loaded with 7.2~tonnes of scintillator, containing $ 3\times 10^{29} $ $^{12}$C
nuclei. The upper limit that can be obtained using formula (8) with these data (with $S_{lim}$=5.91
events for 90\%~c.l. and detection efficiency $ \Delta \varepsilon =0.23 $) is
$\tau (^{12}$C$\rightarrow ^{12}\widetilde{N}+e^{-}+\overline{\nu} )\geq  9.5\cdot10^{27}$ y (90\%~c.l.).
We did
 not cite this approximate result in table 2, because the exact calculation requires precise
knowledge of efficiency of the LSD detector.

\subsection{Limits on non-Paulian $ \beta ^{+}$transitions in $ ^{12}$C: $ ^{12}$C$\rightarrow
^{12}\widetilde{B}+e^{+}+\nu $.}                                     %%%%% BETA++
%-----------------------------------------------------------------------------------------------
The energy release for this reaction, $ Q=17.8 $ MeV, was calculated by assuming that the binding
energy of the neutron in $ S_{1/2} $- shell in $ ^{12}$B nuclei is close to the experimentally
found value for the nuclei $^{11}$B: $ E_{n}(S_{1/2},^{12}B)\simeq E_{n}(S_{1/2},^{11}B)=34.5 $~MeV.
The end-point energy of the $ \beta ^{+} $ spectrum is 16.8~MeV, but the spectrum is shifted
towards higher energies by $ \simeq 0.8 $~MeV due to the registering of annihilation quanta. The
efficiency of the $ ^{12}$C$\rightarrow ^{12}\widetilde{B}+e^{+}+\nu $ transition detection with
energy release $ E>4.5 $~MeV is $ \varepsilon _{\Delta E}=0.31 $. The lower limit on the lifetime
of the proton in the $^{12}$C nuclei is then
\begin{equation}
\tau (^{12}C\rightarrow ^{12}\widetilde{B}+e^{+}+\nu )\geq 7.7\cdot 10^{27}\: y\: (90\%\: c.l.)
\end{equation}
The limits obtained by the NEMO collaboration for this reaction are 3 orders of magnitude weaker: $
\tau (^{12}C\rightarrow ^{12}\widetilde{B}+e^{+}+\nu )\geq 2.6\cdot 10^{24}\: y\: (90\%\: c.l.) $
\cite{NEMO}.

The final results for different PEP violation transitions are shown in table~2 in comparison with
previous results.

%\onecolumn
\begin{table}[!htbp]
\begin{center}
\caption{Mean lifetime limits, $\tau_{\lim}$ (at 90\% C.L.), for non-Paulian transitions in the
CTF2. $E_{0}$ is the average energy of particles, or end-point energy in the case of
$\beta^{\pm}$-transitions; $\triangle E$ is the energy window of CTF2 in which decays were searched
for; $\varepsilon_{\triangle E}$ is the detection efficiency; $S_{\lim}$ the excluded number of
events in the CTF2 spectrum.}
\begin{tabular}{|l|c|c|c|c|c|c|}
  \hline
    Channel & $E_0$, & $\triangle E$ & $\varepsilon_{\triangle E}$& $S_{lim}$ & $\tau_{lim}$ (y) & Previous\\
             & (MeV)       & (MeV) &                   &            &         90\% c.l.         & limits \\ \hline
   $^{12}C\rightarrow^{12}\widetilde{C}+\gamma$ & 17.5 & $\geq$ 4.5 & 4.3$\cdot10^{-2}$ & 2.44 & 2.1$\cdot10^{27}$ & 4.2$\cdot10^{24}$\cite{NEMO}  \\
   $^{16}O\rightarrow^{16}\widetilde{O}+\gamma$ & 21.8 & $\geq$ 4.5 & 5.6$\cdot10^{-3}$ & 2.44 & 2.1$\cdot10^{27}$ & 2.7$\cdot10^{27}$ \cite{KAMIOKANDE}\\
   $^{12}C\rightarrow^{11}\widetilde{B}+ p$     & 4.8-8.2 &  2.0-4.7 & 0.68 x 0.8              & 130 & 5.0$\cdot10^{26}$ & 1.7$\cdot10^{25}$\cite{Ejiri}  \\
   $^{12}C(^{16}O)\rightarrow^{11}\widetilde{C}(^{15}\widetilde{O})+ n$     & 2.2  & 1.7-2.5    & 0.8              & 260  & 3.7$\cdot10^{26}$ & 1.0$\cdot10^{20}$\cite{Kishimoto}  \\
   $^{12}C\rightarrow^{8}\widetilde{Be}+\alpha$ & 2.0  &0.07-0.23& 0.68               & 3300 & 6.1$\cdot$10$^{23}$ & -                  \\
   $^{12}C\rightarrow^{12}\widetilde{N}+ e^{-}+\overline{\nu_{e}}$ & 18.9 & $\geq$ 4.5  & 0.31   & 2.44 & 7.6$\cdot10^{27}$ & 3.1$\cdot10^{24}$ \cite{NEMO}\\
   $^{12}C\rightarrow^{12}\widetilde{B}+ e^{+}+\nu_{e}$ & 17.8 & $\geq$ 4.5  & 0.31   & 2.44 & 7.7$\cdot10^{27}$ & 2.6$\cdot10^{24}$\cite{NEMO} \\
   \hline
 \end{tabular}
\end{center}
\end{table}

\section{Conclusions}
%%%%%%%%%%%%%%%%%%%%%%%%%%%%%%%%%%%%%%%%%%%%%%%%%%%%%%%%%%%%%%%%%%%%%%%%%%%%%%%%%%%%%%%%%%%%%%%
Using the unique features of the Borexino Counting Test Facility -- the extremely low background,
large scintillator mass of 4.2 tonnes, carefully designed muon-veto system and low energy threshold
-- new limits on non-Paulian transitions of nucleons from the $P$-shell to the $1S_{1/2}$-shell in
$^{12}$C and $^{16}$O with the emission of $\gamma, n, p, \alpha$ and $\beta^{\pm}$ particles have
been obtained:

\noindent $\tau(^{12}$C$\rightarrow^{12}\widetilde{C}+\gamma) > 2.1\cdot10^{27}$~y,

\noindent $\tau(^{16}$O$\rightarrow^{16}\widetilde{O}+\gamma) > 2.1\cdot10^{27}$~y,

\noindent $\tau(^{12}$C$\rightarrow^{11}\widetilde{B}+ p)> 5.0\cdot10^{26}$~y,

\noindent
 $\tau(^{12}$C$(^{16}$O$)\rightarrow^{11}\widetilde{C}(^{15}\widetilde{O})+ n)> 3.7 \cdot10^{26}$~y,

\noindent $\tau(^{12}$C$\rightarrow^{8}\widetilde{Be}+\alpha) > 6.1\cdot10^{23}$~y,

\noindent $\tau(^{12}$C$\rightarrow^{12}\widetilde{N}+ e^- + \nu)>7.6\cdot10^{27}$~y

\noindent and

\noindent $\tau(^{12}C\rightarrow^{12}\widetilde{B}+ e^+ + \overline{\nu})> 7.7\cdot10^{27}$ y, all
with 90\% C.L.

Comparing these values with the data of table 2, one can see that these limits for non-Paulian
transitions in $^{12}$C with $\gamma$-,~$p$-,~$n$-,~$\alpha$-, and $\beta^{\pm}$- emissions are the
best to date. The limits on the $\beta^{\pm}$ non-Paulian transitions in $^{12}$C are comparable to
those that can be obtained with the data of the LSD detector \cite{Kekez},\cite{LSD} and the limit
on non-Paulian transition in $^{16}$O with $\gamma$ emission is comparable to the result obtained
using Kamiokande data \cite{KAMIOKANDE}.

\end{document}